\documentclass[
aps, 
pra,
twocolumn,
reprint,
noeprint,
superscriptaddress,
]{revtex4-2}

\usepackage{amsmath}
\usepackage{amssymb}
\usepackage{graphicx}
\usepackage{bm}
\usepackage{color}
\usepackage{mathrsfs}
\usepackage[colorlinks, 
            linkcolor=blue, 
            anchorcolor=blue, 
            citecolor=blue,
            urlcolor=blue
            ]{hyperref}
\usepackage[normalem]{ulem}
\usepackage{orcidlink}

\begin{document}

\title{Efficient generation of multiqubit entanglement states using rapid adiabatic passage}

\author{Shijie Xu~\orcidlink{0009-0005-8782-2457}}
\affiliation{College of Physics, Nanjing University of Aeronautics and Astronautics, Nanjing 211106, China}
\affiliation{Key Laboratory of Aerospace Information Materials and Physics (NUAA), MIIT, Nanjing 211106, China}

\author{Xinwei Li~\orcidlink{0009-0005-9677-0011}}
\email{xinwei.li1991@outlook.com}
\affiliation{Beijing Academy of Quantum Information Sciences, Xibeiwang East Road, Beijing 100193, China}

\author{Xiangliang Li~\orcidlink{0000-0002-7251-7346}}
\email{lixl@baqis.ac.cn}
\affiliation{Beijing Academy of Quantum Information Sciences, Xibeiwang East Road, Beijing 100193, China}

\author{Jinbin Li}
\affiliation{College of Physics, Nanjing University of Aeronautics and Astronautics, Nanjing 211106, China}
\affiliation{Key Laboratory of Aerospace Information Materials and Physics (NUAA), MIIT, Nanjing 211106, China}

\author{Ming Xue~\orcidlink{0000-0002-6156-7305}}
\email{mxue@nuaa.edu.cn}
\affiliation{College of Physics, Nanjing University of Aeronautics and Astronautics, Nanjing 211106, China}
\affiliation{Key Laboratory of Aerospace Information Materials and Physics (NUAA), MIIT, Nanjing 211106, China}

\date{\today}

\begin{abstract}
We propose the implementation of a rapid adiabatic passage (RAP) scheme to generate entanglement in Rydberg atom-array systems. 
This method transforms a product state in a multi-qubit system into an entangled state with high fidelity and robustness. 
By employing global and continuous driving laser fields, we demonstrate the generation of two-qubit Bell state and three-qubit $W$ state,
via sequential RAP pulses within the Rydberg blockade regime.
As an illustrative example, applying this technique to alkali atoms, 
we predict fidelities exceeding 0.9995 for two-qubit Bell and three-qubit $W$ state, along with excellent robustness. 
Furthermore, our scheme can be extended to generate entanglement between weakly coupled atoms and to create four-qubit Greenberger-Horne-Zeilinger states through spatial correlations.
Our approach holds the potential for extension to larger atomic arrays, offering a straightforward and efficient method to generate high-fidelity entangled states in neutral atom systems.
\end{abstract}

\maketitle

\section{Introduction}
Quantum entanglement is a key resource widely used 
in quantum computation~\cite{Henriet2020quantumcomputing,Evered2023High,DiVincenzo2000Computation,Morgado2021Quantum}, 
communication~\cite{Huie2021Multiplexed,Cirac1997Quantum,Stephanie2018Quantum,Kimble2008internet}, 
and metrological tasks~\cite{Degen2017Quantum,aasi2013enhanced,Marciniak2022Optimal}.
The ability to efficiently generate and manipulate high-fidelity entangled states in a quantum system is therefore of paramount importance. 
In recent years, a variety of experimental and theoretical methods have been proposed 
to generate entangled states across different quantum platforms~\cite{Jaksch2000Fast,arute2019quantum, Madjarov19an, Blais21RMP,Ma22PRX,Kim2022,franke2023quantum,bornet2023scalable,Zuo24Entangling}.
Of particular interest are tweezer arrays of neutral atoms, typically separated by micrometer distances~\cite{Barredo16Science}. 
While these atoms do not interact in their ground states, 
they exhibit pronounced interactions upon excitation to high Rydberg states. 
This interaction leads to the Rydberg blockade mechanism, 
where the excitation of one atom prevents nearby atoms from being excited to the Rydberg state within the blockade radius~\cite{Saffman10rmp}.

Many theoretical proposals~\cite{Muller2011,Yang19Dissipative,saffman2020symmetric,Shi2022Quantum,Chang2023High,Dalal23two} 
and experimental implementations~\cite{Wilk2010Entanglement,levine2019parallel,Picken2019Entanglement} 
leveraging the Rydberg blockade mechanism have significantly advanced the field of quantum information processing. 
These studies mainly utilize alkali isotopes such as $^{87}$Rb~\cite{levine2019parallel,Omran2019Generation,Evered2023High,Hines23Spin} 
and $^{133}$Cs~\cite{Graham19Rydberg,Picken2019Entanglement,McDonnel22Demonstration,Graham23midcircuit} 
for their selective excitation to Rydberg states.
Moreover, optical atomic clock transitions in alkaline-earth atoms, 
such as Sr~\cite{Norcia19Science, Schine2022Long, eckner2023realizing} 
and Yb~\cite{Jenkins22PRX, Norcia23midcircuit, Ma2023},
enable the integration of programmable atom arrays with advancements in quantum information processing. 
This integration presents a promising pathway to harnessing highly entangled quantum states 
for entanglement-enhanced quantum metrology~\cite{cao2024multiqubit,finkelstein2024universal,Dominik24Quantum}.

In this work, we propose a rapid adiabatic passage (RAP) scheme for generating multi-qubit entangled states in Rydberg atom-array systems. 
Specifically, we focus on generating two-qubit Bell states, three-qubit $W$ states, and four-qubit Greenberger-Horne-Zeilinger (GHZ)-type states.
These states are characterized by non-separability and cannot be transformed into each other through local quantum operations~\cite{Dur99Separability, Dur00Three, Shao23PRApp}.
GHZ states are maximally entangled and find applications in quantum network and quantum metrology~\cite{Huelga97Improvement,giovannetti2011advances,cao2024multiqubit}, 
while $W$ states are essential for secure quantum communication~\cite{Wang07Quantum,LIU20113160,Lipinska18Anonymous,Miguel-Ramiro_2020,Miguel23Quantum} 
and quantum memory tasks~\cite{Fleischhauer02Quantum,Huie2021Multiplexed}. 
The ability to generate these states on demand is crucial for leveraging their potential in practical quantum information and metrological applications.

Our proposed RAP scheme employs continuous global laser fields to achieve coherent population inversion~\cite{RANGELOV2010Rapid}, 
akin to direct $\pi$-pulse transitions to Rydberg states~\cite{Beterov2011Deterministic, Picken2019Entanglement}, 
thus enabling precise entanglement operations post parameter optimization.
Demonstrated with alkali-atom parameters, 
our approach predicts high fidelities surpassing 0.9995 for two-qubit Bell states and three-qubit $W$ states, 
exhibiting robustness against variations in Rabi frequency and laser frequency detuning.
Additionally, the parameters of the RAP pulse can be directly applied without optimization to larger qubit configurations, achieving high fidelity as well.

Our investigation of the single-excitation property of the RAP pulse distinguishes itself from previous studies~\cite{Beterov2011Deterministic, Beterov2020Application}, 
which primarily focused on multiple atoms in a single trap assuming perfect blockade. 
Unlike these studies, we systematically explore this property under realistic experimental conditions, 
specifically considering the impact of dissipation.
Furthermore, we extend our method to establish entanglement between weakly coupled atoms 
and construct four-qubit GHZ states through selective spatially correlated excitations.
Additionally, our sequential RAP pulse features a distinct pulse shape characterized by continuous frequency detuning
and implementation strategy compared to existing approaches.

\begin{figure}
  \centering
  \includegraphics[width=1\columnwidth]{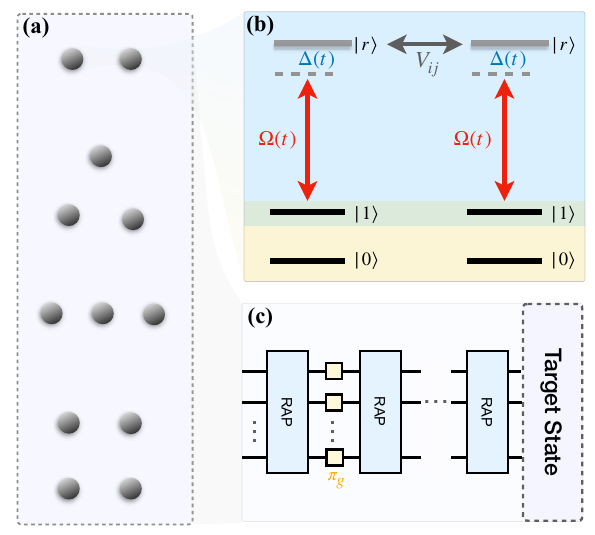}
  \caption{(a) Schematic of possible atom arranging configurations considered in the main text.
  (b) Atomic level structure of two neutral-atom qubits.
  The qubit is encoded in $|0\rangle$ and $|1\rangle$, with a high-lying Rydberg state $|r\rangle$.
  A global laser dresses only the ground state $|1\rangle$, coupling it to $|r\rangle$, with a Rabi frequency $\Omega$ (red) and detuning $\Delta$ (blue).
  A $\pi_g$ pulse can be applied in the ground state manifold between $|0\rangle$ and $|1\rangle$ (yellow shaded area).
  (c) Scheme to generate target entangled states realized by sequential RAP pulses with inserted ground state $\pi_g$ pulse (yellow square).  
  }
\label{fig:scheme}
\end{figure}

This paper is organized as follows. Section~\ref{sec:model} introduces the model Hamiltonian and our RAP scheme. 
In Section~\ref{Generation of Multi-Qubit W States}, 
we present the scheme for generating two-qubit Bell states and three-qubit $W$ states, 
and then in Section~\ref{Analysis of Fidelity and Robustness}, we analyze their fidelity and robustness.
Sections~\ref{Achieving Entanglement of Weakly Coupled Atoms} 
and \ref{GHZ State Entanglement in Atomic Arrays} propose schemes for establishing entanglement between weakly coupled atoms 
and discuss the generation of GHZ states. 
Finally, in Section~\ref{Conclusions}, we conclude with a discussion.

\begin{figure*}
  \centering
  \includegraphics[width=1.99\columnwidth]{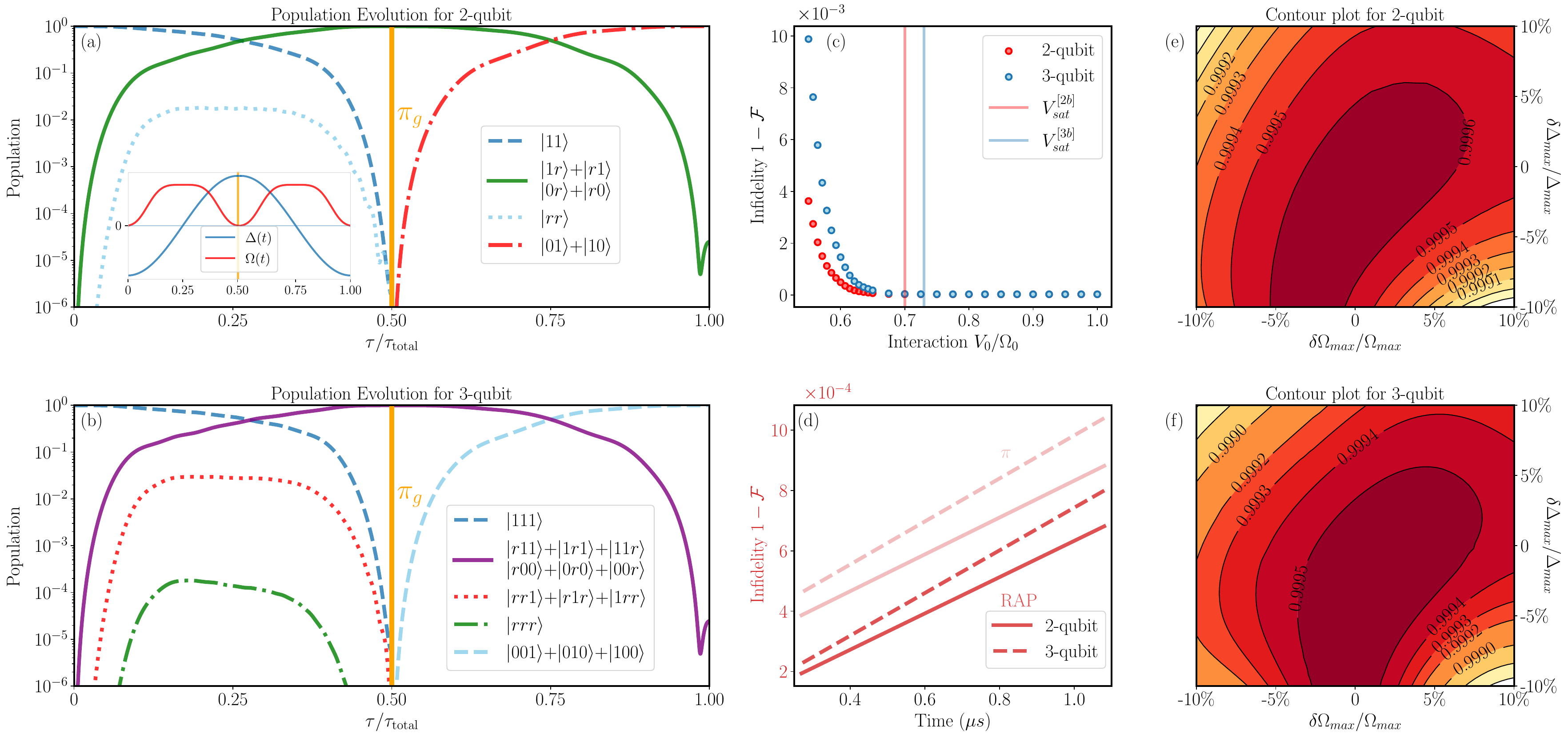}
  \caption{ Multiqubit $W$ states generation protocol.
  (a) State evolution process for 2-qubit with \(V_0=0.70\Omega_0\). The inset illustrates the pulse and detuning shapes as determined by Eq.~\eqref{eq:scheme}.
  (b) State evolution process for 3-qubit with $V_0=0.73\Omega_0$.
  (c) Infidelity (without dissipation) as a function of interaction strength. 
  Red and blue vertical lines indicate the chosen saturation interactions that maximize 2- and 3-qubit fidelity, respectively.
  (d) Infidelity as a function of total time, achieved by varying \(\Omega_0\). 
  Deep red represents the RAP scheme, while the light line represents \(\pi\) pulses replacing the RAPs. Solid lines are for 2-qubit, dashed lines for 3-qubit.
  (e) Contour plot of fidelity with parameter variation, setting $\Omega_0/(2\pi)=100$ MHz and $V_0/(2\pi)=70$ MHz for 2-qubit.
  (f) Contour plot of fidelity with parameter variation, setting $\Omega_0/(2\pi)=100$ MHz and $V_0/(2\pi)=73$ MHz for 3-qubit.  
  }
  \label{fig:two qubit scheme}
\end{figure*}

\section{Model and scheme}\label{sec:model}
The proposed scheme involves a three-level system, as depicted in Fig.~\ref{fig:scheme}(b). 
The atom features two ground states, $|0\rangle$ and $|1\rangle$, along with a high-lying exicted Rydberg state $|r\rangle$. 
A pulse couples the $|1\rangle$ state to the $|r\rangle$ state with Rabi frequency $\Omega$ and laser frequency detuning $\Delta$. 
Due to the global laser dressing, all atoms experience identical laser detunings and Rabi frequencies.
The Hamiltonian of an $n$-qubit system is given by:
\begin{equation}\label{eq:Ht} 
  \hat{H}(t) = \Delta(t) \sum_{i=1}^n |r\rangle_i \langle r|_i 
  + \frac{\Omega(t)}{2} \sum_{i=1}^n (|r\rangle_i \langle 1|_i+ {\rm H.c.}) + \hat{V}_{\rm RRI},
\end{equation}
where $\hat{V}_{\rm RRI} = \sum_{i<j}^n V_{ij} |r\rangle_i \langle r|_i \otimes |r\rangle_j\langle r|_j$ 
is the Rydberg-Rydberg interaction between each pair of atoms, 
with strength $V_{ij}$ strongly depending on the distance between the atoms and the specific type of Rydberg states~\cite{Singer_2005}.
Furthermore, a fast $\pi$-pulse, denoted as $\pi_g$, 
can be applied in the ground state manifold through stimulated Raman transitions or microwave driving 
to coherently invert the population between $|0\rangle$ and $|1\rangle$
[yellow shaded area in Fig.~\ref{fig:scheme}(b)],
in contrast to the coupling between $|1\rangle$ and $|r\rangle$ [blue shaded area in Fig.~\ref{fig:scheme}(b)].

Figure~\ref{fig:scheme}(c) presents a circuit representation of our entanglement approach, 
which employs repeated RAP pulses and a $\pi_g$-pulse between RAPs to facilitate entanglement generation.
Our scheme is inspired by Refs.~\cite{Picken2019Entanglement,Wilk2010Entanglement}
but utilizes RAP pulses to achieve coherent population inversion.
We define $\tau = \Omega_0 t$ as the dimensionless time, 
with $\Omega_0$ as the unit of Rabi frequency, where $\tau\in [0,\tau_{\rm tot}]$. 
The Rabi frequency $\Omega(\tau)$ and laser detuning $\Delta(\tau)$ are set as follows\,\cite{saffman2020symmetric}:
\begin{align}\label{eq:scheme}
    \Omega(\tau) &= \Omega_{\text{max}} (e^{-[(\tau-\tau_k)/\tau_R]^4} - a)/(1 - a) ,\nonumber\\
    \Delta(\tau) &= (-1)^{k+1} \Delta_{\text{max}} \sin(\pi (\tau-\tau_k)/\tau_D),
\end{align}
for $(k-1)/2 \leq \tau/\tau_{\rm tot} \leq k/2$ with $\tau_k=(2k-1)\tau_{\rm tot} /4$ and $k \in \{1,2,\dots\}$, 
representing the $k$-th RAP pulse. 
We set $\tau_{R}=0.175\tau_{\rm tot} $ and $a=\exp[-(\tau_{\rm tot} /4\tau_R)^4]$
such that $\Omega(\tau)$ is zero at the beginning and end of each sweep process.
Additionally, $\tau_D=\tau_{\rm tot} /2$ ensures that the detuning profile is continuous and 
completes  a full sinusoidal cycle within the time $\tau_{\rm tot}$.
To provide a visual representation of the scheme described in Eq.~\eqref{eq:scheme}, we have illustrated the pulse and detuning shapes in the 
inset of Fig.~\ref{fig:two qubit scheme}(a).
The time required for the \(\pi_g\)-pulse (in \cite{Madjarov2020}, this time is approximately 70 ns) 
between the two RAP pulses is considered negligible compared to the time scale of 0.5-1.0 \(\mu\)s in our scheme. 
The RAP pulse widths  $\{\tau_R, \tau_D\}$ and amplitudes $\{\Delta_{\rm max}, \Omega_{\rm max}\}$ 
will be optimized for pulse shaping,
with the figure of merit being the fidelity, defined as~\cite{Pelegrí_2022}:
\begin{equation}
  \mathcal{F} = |\langle \psi _{\text{tar}}|\hat{\rho}_f|\psi _{\text{tar}}\rangle|^2,
\end{equation}
where $|\psi_{\text{tar}}\rangle$ is the target multiqubit entangled state, 
and $\hat{\rho}_f$ is the density operator of the state evolved to $\tau_{\rm tot}$ 
under the Hamiltonian $H(t)$ in Eq.~(\ref{eq:scheme}),
and the Lindblad superoperator $\mathcal{L}$ to account for possible dissipative effects.
Thus, the numerical simulation is governed by the master equation,
\begin{equation}\label{eq:mastereq}
\partial_t \hat{\rho}(t) = i[\hat{\rho}, \hat{H}(t)] + \mathcal{L}[\hat{\rho}],
\end{equation}
where $\mathcal{L}[\hat{\rho}] = \sum_k \mathcal{D}_k[\hat{\rho}]$, 
and the dissipators 
$$\mathcal{D}_k[\hat{\rho}] = \hat{L}_k \hat{\rho} \hat{L}_k^\dagger - \frac{1}{2}(\hat{L}_k^\dagger \hat{L}_k \hat{\rho} + \hat{\rho} \hat{L}_k^\dagger \hat{L}_k)$$ 
describe the dissipative processes that induce fidelity losses. 
Optimizing the parameters $\{\Delta_{\rm max}, \Omega_{\rm max}, \tau_R, \tau_D\}$ in ${H}(t)$ is crucial to achieving high fidelity in the entangled state preparation.

In the following sections, we present two distinct scenarios.
The first scenario involves a symmetric configuration with respect to the Rydberg blockade, 
where atoms are paired within the blockade radius. 
This setup facilitates the generation of entangled states using a straightforward RAP process, 
requiring only two iterations. The laser frequency detunings in Eq.~\eqref{eq:scheme} are kept continuous throughout the process. 
Notably, the optimized parameters for this scheme are universal and applicable across various small qubit configurations, 
eliminating the need for individual parameter optimization.

In contrast, the second scenario presents that not all atoms are situated within the Rydberg blockade radius. 
In this case, some atom pairs fall within the blockade regime, while others are situated outside. 
To mitigate this heterogeneity, an auxiliary relay atom is introduced between atom pairs with weak interaction. 
Alternatively, the spatial arrangement facilitates tailored atom excitations, 
enabling the generation of high-fidelity and robust entangling operation.

\section{Inside the Blockade regime}
\label{sec:Wstate1Excitation}

\subsection{Two-qubit Bell and three-qubit W states}
\label{Generation of Multi-Qubit W States}

To generate a two-qubit Bell state, we start with the two atoms in state $|11\rangle$ 
with strong Rydberg-Rydberg interaction strength $V_0$ within the blockade regime.
We adopt the profiles in Eq.~(\ref{eq:scheme}) for the RAP pulse, 
with optimized parameters $\Omega_{\text{max}} = 0.17\Omega_0$ and $\Delta_{\text{max}} = 0.24\Omega_0$ to achieve the optimal population transfer.
The first RAP results in a transfer from the $|11\rangle$ state to the $\frac{1}{\sqrt{2}} (|1r\rangle + |r1\rangle)$ state due to the Rydberg blockade.
Subsequently, there are two approaches to achieve a two-qubit Bell state encoded in ground states $|0\rangle$ and $|1\rangle$:
(i) Apply a fast $\pi_g$ pulse between $|0\rangle$ and $|1\rangle$, and then followed by another RAP sweep, to generate the Bell state $\frac{1}{\sqrt{2}}(|10\rangle + |01\rangle)$\,\cite{Picken2019Entanglement}. 
The overall population evolution of the states can be observed in Fig.~\ref{fig:two qubit scheme}(a). 
It is evident that the state transfer is almost perfect, indicating that the desired state is fully transferred during the RAP process. 
(ii) Alternatively, adjust the laser to couple states $|0\rangle$ with $|r\rangle$, 
while decoupling $|1\rangle$ and $|r\rangle$, 
and then perform an RAP sweep to achieve the transition $\frac{1}{\sqrt{2}} (|1r\rangle + |r1\rangle)$ $\to$ $\frac{1}{\sqrt{2}} (|10\rangle + |01\rangle)$~\cite{Wilk2010Entanglement}.
If the two ground states are taken from hyperfine levels with small energy splitting, 
specific configurations of laser polarization are needed, 
or one can encode the qubit in fine-structure levels~\cite{pucher24fine,Unnikrishnan24coherent}.
Both approaches can achieve the Bell state $\frac{1}{\sqrt{2}} (|10\rangle + |01\rangle)$ using double RAP pulses. 
Assuming the time required for the $\pi_g$ pulse and laser frequency changes are negligible, 
the achieved Bell state fidelity of both approaches remains almost the same.

Now we extend to three qubits.
To achieve optimal single Rydberg excitation symmetry within the blockade regime,
three atoms are placed at the vertices of an equilateral triangle~\cite{Hou2019Wstate, Pelegrí_2022}.
The pulse sequence for three qubits is the same as in the two-qubit case,
using identical parameters and involving two RAP pulses with a $\pi_g$ pulse sandwiched in between.
The state evolves starting from $|111\rangle$ as follows:
A single RAP sweep realizes a state transition to $\frac{1}{\sqrt{3}}(|11r\rangle + |1r1\rangle + |r11\rangle)$. 
This is followed by a fast $\pi_g$ pulse between the ground states $|0\rangle$ and $|1\rangle$.
Another RAP pulse is then performed, eventually preparing the $W$ state $\frac{1}{\sqrt{3}} (|001\rangle + |010\rangle + |100\rangle)$. 
The state evolution process of three qubits is shown in Fig.~\ref{fig:two qubit scheme}(b). 
It can be observed that the single-excitation property is well maintained, and the double-excitation component remains small due to the Rydberg blockade.
The proposed scheme is adaptable to more qubits. 
Intuitively, a three-dimensional atomic arrangement might be essential to ensure single excitation across multiple atoms. 
For instance, with four qubits, a three-dimensional equilateral pyramid would yield optimal single-excitation performance. 
However, arranging atoms in three dimensions is more challenging than in two dimensions in practical experiments~\cite{Barredo16Science,Barredo18Synthetic,Schymik20Enhanced}.

\subsection{Fidelity and robustness}\label{Analysis of Fidelity and Robustness}
To incorporate dissipative effects, we assess fidelity in a system of neutral Cs atoms 
utilizing $|r\rangle=|107P_{3/2}; m_J = 3/2\rangle$ as the Rydberg state~\cite{saffman2020symmetric}, 
with dissipation channels $\hat{L}_1=\sqrt{\gamma_r/16}|0\rangle \langle r|$, 
$\hat{L}_2=\sqrt{\gamma_r/16}|1\rangle \langle r|$, 
and $\hat{L}_3=\sqrt{7\gamma_r/8}|r\rangle \langle r|$ in Eq.~(\ref{eq:mastereq}). 
The total dissipation rate is $\gamma_r = 1/(540 \,\mu s)$
~\cite{saffman2020symmetric}.
In Fig.~\ref{fig:two qubit scheme}(c), we find that increasing the interaction strength does not notably improve fidelity, 
signifying the onset of saturation interaction, henceforth referred to as $V_{\rm sat}$,
as demonstrated by vertical solid lines.
We then optimize the parameters under realistic $\Omega_0$, 
including the maximal Rabi frequency $\Omega_{\text{max}}$, maximal detuning $\Delta_{\text{max}}$,
and also the saturation interaction strength $V_{\rm sat}$. 
In Fig.~\ref{fig:two qubit scheme}(c), we determined that to achieve optimal fidelity, 
interaction strengths of $V_0 \geq V^{\rm [2b]}_{\rm sat}\simeq 0.70\Omega_0$ 
for the 2-qubit system, and $V_0 \geq V^{\rm [3b]}_{\rm sat}\simeq 0.73\Omega_0$ for the 3-qubit setup are adequate, 
without considering dissipative effects.

One can use also resonant $\pi$-pulse between the $|1\rangle$ and $|r\rangle$ states to achieve coherent population inversion, 
similar to the aforementioned RAP pulses.
However, the Rabi frequency will differ for different numbers of qubits~\cite{levine2019parallel,Pelegrí_2022}, 
and population in the Rydberg states will adversely affect the fidelity due to dissipative effects. 
We have compared the fidelity of our RAP scheme (light solid red line) with that of the 
$\pi$-pulse scheme (deep red solid line) over different total times, as shown in Fig.~\ref{fig:two qubit scheme}(d).
It is evident that the fidelity of the RAP scheme consistently outperforms the 
$\pi$-pulse scheme for transitions between $|1\rangle$ and $|r\rangle$ scheme.
With increasing total evolution time $\tau_{\rm tot}$, 
the fidelity decreases due to spontaneous emission and dephasing of the Rydberg state, 
but still maintains over 0.9993 within 1\,$\mu$s. 

\begin{figure}
  \centering
  \includegraphics[width=.97\columnwidth]{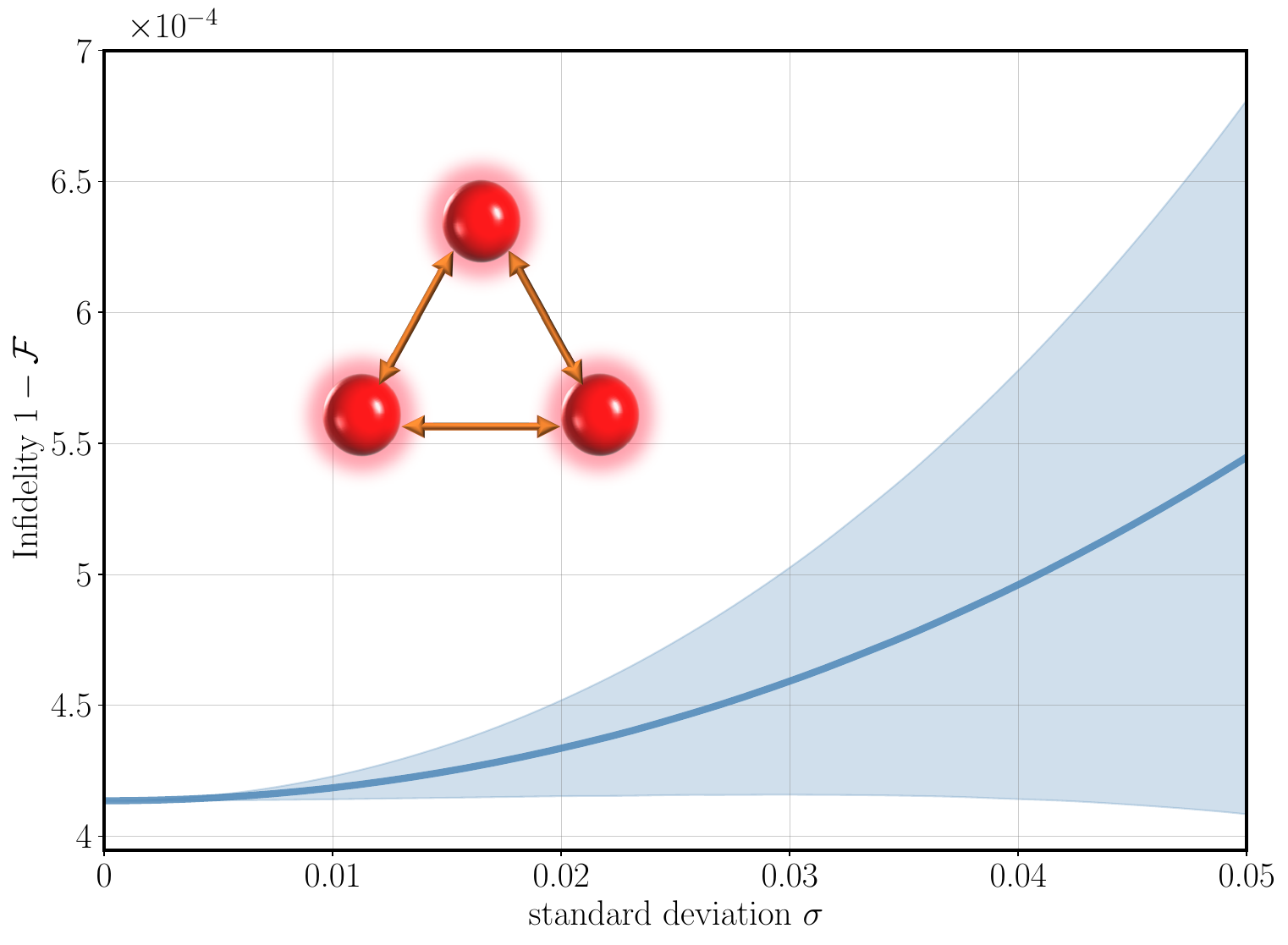}
  \caption{Infidelity variation due to positional fluctuations for the 3-qubit case. 
  The blue solid line represents the mean, while the light blue shading indicates the region within one standard deviation. 
  This illustrates the robustness of our scheme to interaction asymmetries caused by positional fluctuations.}
\label{fig:pos3qubit}
\end{figure}

We analyze the robustness of our present scheme using $\Omega_0/(2\pi) = 100$ MHz for Cs atoms, 
considering Rabi pulse strength and laser detuning fluctuations within $\pm 10\%$. 
As depicted in Fig.~\ref{fig:two qubit scheme}(e) for 2-qubit case 
and Fig.~\ref{fig:two qubit scheme}(f) for 3-qubit case, 
despite such significant fluctuations, the W-state fidelity remains above 0.999 for both cases.
This demonstrates that our proposed scheme is highly robust against parameter variations, 
which is one of the advantages of RAP.
Regarding the impact of Doppler frequency shifts, 
we categorize them as fluctuations in detuning frequency~\cite{saffman2020symmetric, Chang2023High}.
For Cs atoms at a temperature of 10 $\mu$K, 
the detuning fluctuation is only about $\pm0.3\%$~\cite{saffman2020symmetric},
indicating that the impact of Doppler shifts on our scheme is minute.

Furthermore, we evaluate the robustness of our scheme against atomic position fluctuations for the 3-qubit case. 
We start with an initial configuration where atoms occupy the vertices of an equilateral triangle [cf.~inset of Fig.~\ref{fig:pos3qubit}] with a unit side length, 
and an initial interaction of $(2\pi)\,200$\,MHz between adjacent atoms at a distance of 8.14\,$\mu$m~\cite{saffman2020symmetric}.
These atoms are then subjected to position fluctuations, 
modeled as a two-dimensional uncorrelated Gaussian distribution, 
$\mathbf{X}_i\equiv(x_i, y_i)\sim N(0,0, \sigma^2,\sigma^2)$, where $i = 1, 2, 3$. 
For each value of \(\sigma\), we randomly sample the positions of the three vertices according to the Gaussian distribution 
and recalculate the interaction between adjacent atoms.
To assess the impact of these position fluctuations, we calculated the average fidelity 
over 30 random sampled configurations for every $\sigma$. 
As depicted in Fig.~\ref{fig:pos3qubit}, 
the light blue shading represents one standard deviation, 
and the solid line indicates the average fidelity. 
For the maximum considered $\sigma=0.05$, these fluctuations translate into an interatomic distance variation of 
$[-400, 400]$\,nm along two orthogonal coordinates within one standard deviation.
Even with such significant atomic position fluctuations, the fidelity remains above 0.9993, 
underscoring the robustness and potential applicability of our scheme in realistic scenarios.

\begin{figure}
  \centering
  \includegraphics[width=1\columnwidth]{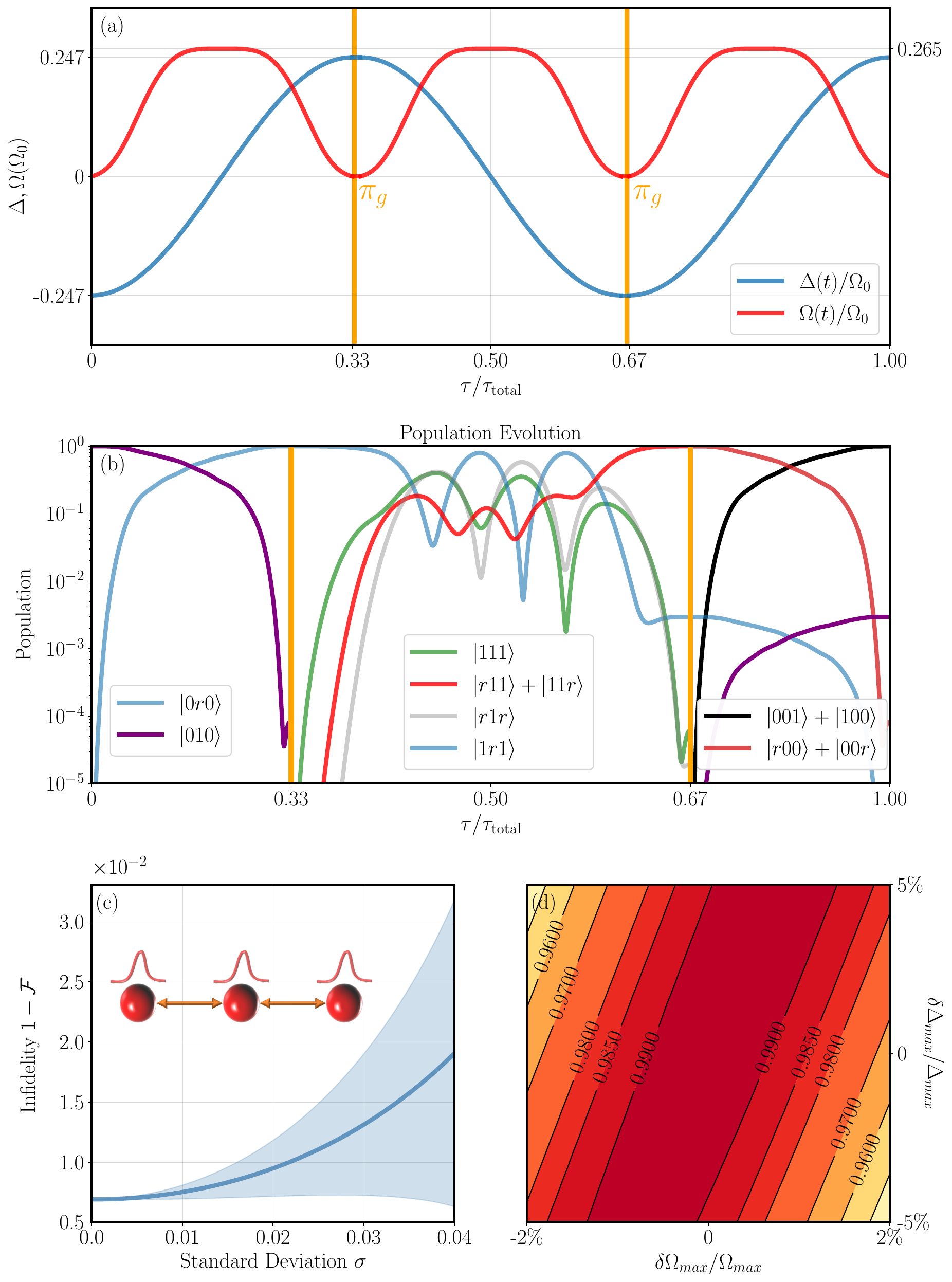}
  \caption{Scheme for generating entanglement among weakly coupled atoms. 
  (a) RAP pulse profile with parameters \(\Delta_{\text{max}} =0.247\Omega_0\), \(\Omega_{\text{max}} =0.265\Omega_0\).
  (b) State evolution process when \(V_0=0.96\Omega_0\), with an enlarged view of the second sweep. 
  (c) Fidelity under positional fluctuations around the original positions, modeled by a Gaussian function with standard deviation \(\sigma\). 
  The dashed blue shaded area indicates the region within one standard deviation, with the solid line representing the mean. 
  (d) Contour plot of the fidelity changing with respect to the variation of parameters $\Delta_{\text{max}}$ and $\Omega_{\text{max}}$, for \(\Omega_0/(2\pi)=100\) MHz and \(V_0/(2\pi)= 96\) MHz.
  }
\label{fig:3qubitscheme}
\end{figure}

For the three-qubit case, we consider a two-dimensional triangular arrangement with symmetric blockade between any two atoms. 
However, for four qubits, three-dimensional configurations are experimentally challenging~\cite{Barredo18Synthetic}, 
so we focus on two-dimensional asymmetric distributions.
Using the same RAP parameters and setting $\Omega_0/(2\pi)=100$ MHz, 
we obtain similar fidelity for four qubits whether the atoms are arranged in a three-dimensional equilateral pyramid or a two-dimensional square, 
assuming sufficiently large interaction strength. 
This results in a fidelity of 0.9997, demonstrating the scheme's tolerance to asymmetry 
and its potential for generating single-excitation entangled $W$-states in large-scale atomic arrays.

\section{Outside the Rydberg blockade regime}
\label{sec:outsideBlockade}
We have discussed the single-excitation properties induced by RAP within the Rydberg blockade regime. 
In this section, we focus on entanglement state preparation achieved by strategically arranging atomic positions 
and leveraging interaction strengths between atom pairs. 
This exploits the Rydberg blockade, with distant atoms outside the blockade radius and adjacent atoms within it, 
enabling specific atomic excitations.

\subsection{Two-qubit and three-qubit entangling operations}
\label{Achieving Entanglement of Weakly Coupled Atoms}

The entangling operation we discussed earlier rely on strong Rydberg interaction. 
Here, we propose generating Bell states between two atoms with weak interaction using an ancillary qubit. 
This involves placing three atoms in a straight line with equal spacing~\cite{Barredo2014Demonstration, Beterov2018Fast}, as shown in Fig.~\ref{fig:3qubitscheme}(c) (inset).
The scheme requires strong interaction between adjacent atoms although nearly zero interaction between the end atoms, 
making it well-suited to the van der Waals interaction, 
which scales as the inverse sixth power of the atomic distance~\cite{Walker2008Consequences,Zwierz2009High}.
This configuration allows us to achieve entanglement generation between distant atoms by strategically moving a single relay atom into position~\cite{Shaw24multiensemble}.

For realistic parameter analysis, we consider the Rydberg state $|r\rangle=|70S_{1/2}; m_J = -1/2\rangle$ of $^{87}$Rb atoms~\cite{levine2019parallel,Keesling2019,Omran2019Generation,Bernien2017}. 
Initially, three atoms are prepared in the state $|010\rangle$, 
with the middle atom initialized in the $|1\rangle$ state. 
The interaction between adjacent atoms is set to $V_0=0.96\Omega_0$, 
resulting in a weaker interaction of $V_1=V_0/2^6=0.015\Omega_0$
between atoms separated by a larger distance. 
We apply three global RAP pulses, as shown in Fig.~\ref{fig:3qubitscheme}(a), 
with a $\pi_g$ pulse inserted between each pair of RAP pulses and continuously changing frequency detuning. 
This setup ensures optimal conditions for entanglement generation, with a total scaled time of 
$\Omega\tau_{\rm tot}/(2\pi)=81$.

The state transformation proceeds as follows:
First, $|010\rangle \xrightarrow[]{\text{RAP}}|0r0\rangle\stackrel{\pi_g}{\longrightarrow}|1r1\rangle$.
Then, the second RAP pulse symmetrically generates the state $\frac{1}{\sqrt{2}}(|11r\rangle + |r11\rangle)$, 
which transforms to $\frac{1}{\sqrt{2}}(|00r\rangle + |r00\rangle)$  after the second $\pi_g$ pulse.
Finally, with the third RAP pulse, we obtain $\frac{1}{\sqrt{2}}(|001\rangle + |100\rangle)$.
The middle atom can then be traced out, yielding the Bell state $\frac{1}{\sqrt{2}}(|01\rangle + |10\rangle)$
between the two initially weakly interacting qubits. 
The atomic population during the evolution is shown in Fig.~\ref{fig:3qubitscheme}(b).
As depicted, the transition to the target state is nearly complete by the end of the process.

We first consider $\Omega_0 = (2\pi)100$ MHz (corresponding to a total time of 0.81 $\mu$s) and a Rydberg state decay rate of 
$\gamma_r=1/(147\mu s)$~\cite{Pritchard2017ARC,Beterov09PRA}.
Additionally, we introduce a dump level \(|d\rangle\), which is designed to capture all decays from the Rydberg state. 
Under these conditions, the scheme achieves a fidelity of 0.993.
To assess its robustness against atomic positional fluctuations, we examined their impact on fidelity.
Starting with atoms separated by a unit distance and interacting with $V_0=(2\pi)\,96$ MHz,
each atom underwent one-dimensional Gaussian fluctuations along the line with amplitude $\sigma$.
For $\sigma=0.04$, this results in a positional variation of $[-230, 230]$\,nm within one standard deviation.
We calculated the average fidelity and standard deviation over 30 random samples for each $\sigma$,
shown in Fig.~\ref{fig:3qubitscheme}(c),
where the light blue shadow represents one standard deviation and the solid line denotes the mean.
We calculated the average fidelity and standard deviation over 30 random samples for each $\sigma$, 
as shown in Fig.~\ref{fig:3qubitscheme}(c). In these tests, the fidelity remains above 0.97, 
demonstrating the robustness of the scheme against positional fluctuations.
We also assessed the robustness of this scheme to optimized parameter fluctuations, 
as illustrated in the contour plot in Fig.~\ref{fig:3qubitscheme}(d). 
Fluctuations of $\pm$5\% in $\Delta_{\text{max}}$ and $\pm$2\% in $\Omega_{\text{max}}$ were considered.
The fidelity remains above 0.980 across most of the parameter space, 
with slight reductions observed at the edges of the plot. 
Note that in Ref.~\cite{Omran2019Generation}, the atom temperature of approximately $10\,\mu$K leads to fluctuating Doppler shifts in the addressing lasers of order $(2\pi)\,43$ kHz, 
as well as fluctuations in atomic position that result in variations in Rydberg interaction strengths, 
which is well below the 0.5\% parameter variation considered here.
Consequently, fidelity remains above 0.990 from Fig.~\ref{fig:3qubitscheme}(d).
This analysis highlights the scheme's greater resilience against variations in laser detuning compared to fluctuations in Rabi frequency. 
Despite these challenges, this scheme provides valuable insights into achieving entanglement among weakly coupled atoms.

\begin{figure}
  \centering
  \includegraphics[width=1.\columnwidth]{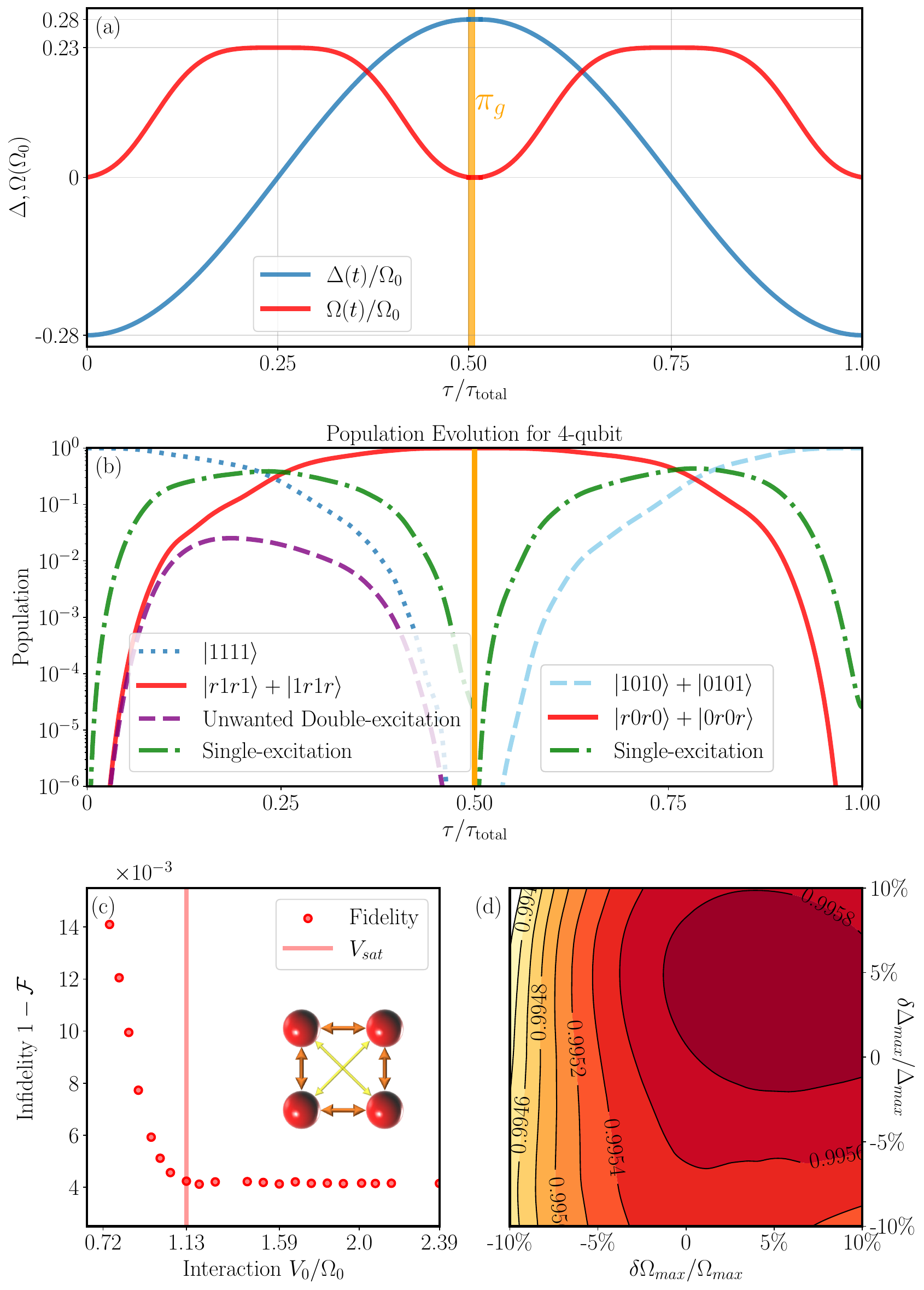}
  \caption{The scheme for generating a four-qubit GHZ state. 
  (a) shows the RAP pulse diagram with parameters \(\Delta_{\text{max}} =0.28\Omega_0\), \(\Omega_{\text{max}} = 0.23\Omega_0\). 
  (b) depicts the state evolution process when $V_0=1.13\Omega_0$, 
  while unwanted double-excitation refers to the double Rydberg-excitation states other than \(|1r1r\rangle + |r1r1\rangle\).
  (c) presents the fidelity (without dissipation) as a function of \(V_0\) under the conditions \(\Delta_{\text{max}} = V_0/4.05\) and \(\Omega_{\text{max}} = V_0/4.85\), then $1.13\Omega_0$ is the situation interaction.
  (d) shows a contour plot for \(V_0=(2\pi)113\) MHz, indicating that the chosen parameters do not yield the optimal fidelity.
  }
\label{fig:4qubitscheme}
\end{figure}

\subsection{The four-qubit GHZ state via double excitations}\label{GHZ State Entanglement in Atomic Arrays}

Furthermore, we demonstrate generating four-qubit GHZ states by positioning four atoms 
at the vertices of a square [cf.~inset of Fig.~\ref{fig:4qubitscheme}(c)]
and utilizing van der Waals interactions to induce specific atomic excitations.
The interaction between adjacent atoms is $V_0$, and for diagonal atoms, it is $V_1 = V_0/\sqrt{2}^6$.
Parameters are set to $V_0/\Delta_{\text{max}} = 4.05$ and $V_0/\Omega_{\text{max}} = 4.85$,
which may need adjustment for varying interaction strengths.
This ensures the adiabatic condition $V_1 \ll \Delta_{\text{max}}, \Omega_{\text{max}} \ll V_0$
holds across different interaction strengths~\cite{Comparat2009General}.
It is worth noting that our linear relationship between parameters $\{\Omega_{\rm max}, \Delta_{\rm max}\}$ and $V_0$ is a simple empirical rule 
and may not maximize fidelity.
As a consequence, the Rydberg blockade leads to singly excited adjacent atoms, 
while weaker interactions result in doubly excited diagonal atoms, 
establishing symmetry conducive to GHZ state construction.
For optimal fidelity, further optimization of $\Delta_{\text{max}}$ and $\Omega_{\text{max}}$ 
is necessary across different $V_0$ values.
The global RAP pulse in Fig.~\ref{fig:4qubitscheme}(a) remains consistent with previous setups, 
with a $\pi_g$ pulse sanwithed by two RAP pulses.
We choose $V_0= 1.13\Omega_0$ as an example, 
the parameter chosen is depicted in Fig.~\ref{fig:4qubitscheme}(a). 
Initially, atoms are prepared in state $|1111\rangle$.
The state transformation proceeds as follows:
\begin{eqnarray}
  |1111\rangle&\xrightarrow[]{\text{RAP}} &\frac{|1r1r\rangle + |r1r1\rangle}{\sqrt{2}}\xrightarrow[]{\pi_g} \frac{|0r0r\rangle + |r0r0\rangle}{\sqrt{2}}\nonumber\\
  &\xrightarrow[]{\text{RAP}}&\frac{1}{\sqrt{2}}(|0101\rangle + |1010\rangle),
\end{eqnarray}
where the specific double Rydberg excitation can be revealed in Fig.~\ref{fig:4qubitscheme}(b) during the full evolution process. 
The fidelity plot in Fig.~\ref{fig:4qubitscheme}(c) shows saturation interaction at $V_0=1.13\Omega_0$.
Considering dissipation and taking $\Omega_0=(2\pi)\,100$ MHz, 
the fidelity achieved for $V_0 = (2\pi)113$ MHz is approximately 0.9956.
Figure~\ref{fig:4qubitscheme}(d) demonstrates the robustness against parameter fluctuations, 
further indicating suboptimal parameter choices for specific $V_0$.
Additionally, a local qubit $\pi_g$ pulse can flip the state of every other site to prepare the canonical form of the
GHZ state, $(|0000\rangle+|1111\rangle)/\sqrt{2}$, from $(|0101\rangle + |1010\rangle)/\sqrt{2}$~\cite{Omran2019Generation}.

Expanding on our approach, which utilizes the relationship between atomic positions and interactions to achieve 
specific atomic excitations for entangled states, 
we foresee extending this method to larger atomic arrays. 
In larger setups, the spatial arrangement facilitates tailored atom excitations~\cite{Shaw24multiensemble}, 
thereby enabling the generation of high-fidelity and robust GHZ states.
This approach aligns with methodologies described in Refs.~\cite{Omran2019Generation,Zwierz2009High},
indicating potential scalability of our scheme.

\section{Conclusions and Discussions}\label{Conclusions}
We have presented a robust and efficient method for generating high-fidelity entangled states in neutral 
atom systems using rapid adiabatic passage (RAP) pulses. 
By leveraging the single-excitation mechanism within the Rydberg blockade radius, 
we have demonstrated the creation of multi-qubit $W$ states with fidelities exceeding 0.9995 for two-qubit and three-qubit systems. 
The proposed scheme has shown excellent robustness against pulse parameter fluctuations and Doppler frequency shifts, 
making it suitable for practical experimental implementations.
Additionally, we have extended our approach to generate specific entangled states by strategically 
positioning atoms inside and outside the Rydberg blockade radius. 
This spatial correlation technique has enabled the creation of entanglement between weakly coupled atoms 
and the generation of four-qubit GHZ states. 
The flexibility and scalability of our method highlight its potential for application 
in larger atomic arrays and more complex quantum information processing tasks, 
including clock-qubit arrays for entanglement-enhanced measurements.

\section*{Acknowledgements}
This work was supported by Innovation Program for Quantum Science and Technology (Grant No.\,2021ZD0302100),
National Natural Science Foundation of China (Grant No.\,12304543 and No.\,92265205), 
and Open Research Fund Program of the State Key Laboratory of Low-Dimensional Quantum Physics (KF202111).
M.X. was also supported by the Innovative and Entrepreneurial Talents Project of Jiangsu Province (Grant No. JSSCBS20220234) 
and the Changzhou Longcheng Talent Plan-Young Science and Technology Talent Support Project.
S.X. was supported by Innovation Training Program of NUAA (No.\,20241007700100Z and No.\,2024CX021030).

\bibliography{mybib}

\end{document}